\documentclass[amssymb,prb,aps,showpacs,twocolumn,showpacs]{revtex4}
\usepackage{graphicx}
\begin{document}

\title{\textbf{Green's-function method for calculation of  adsorption of organic
 molecules on noble metal nanoparticles}}
\author{O.V.Farberovich$^{1,2}$, B.D.Fainberg$^{1,3}$,
V.G.Maslov$^{4}$, V.Fleurov$^{2}$}
\affiliation{
$^{1}$ Faculty of Sciences, Holon Institute of Technology,\\
52 Golomb Street, Holon   58102, Israel\\
$^{2}$  Raymond and Beverly Sackler Faculty of Exact Sciences,\\
School of Physics and Astronomy, Tel-Aviv University, Tel-Aviv 69978, Israel\\
$^{3}$  Raymond and Beverly Sackler Faculty of Exact Sciences,\\
School of Chemistry, Tel-Aviv University, Tel-Aviv 69978, Israel\\
$^{4}$ Saint-Petersburg State University of Information and Technologies,\\
Mechanics and Optics, Saint-Petersburg 197101, Russia}
\begin{abstract}
A numerical method for the calculation of electronic structure of a
nanosystem composed of a pseudoisocyanine (PIC) molecule assembled
on a silver nanoparticle is developed. The electronic structure of
the silver nanoparticle containing 125 atoms is calculated within
the local density version of the density functional method. A model
of an Ag atom embedded in the center of a spherical jellium cluster
is used. The host electron Green's function is calculated by means
of the spherically symmetric expansion. The principal theoretical
tool is the scattering theory using the Green's-function method. The
molecule -- silver nanosystem interaction is studied using the
approach similar to that of the Anderson model for transition metal
impurities in solids. Localized levels are shown to split off from
the top of the band of the Ag nanoparticle. The electronic structure
calculations yield information on the character of chemical bonding
in the PIC molecule --- silver particle nanosystem.
\end{abstract}
\maketitle
\section{Introduction}

Organic-inorganic nanohybrid materials that utilize noble metals
(silver or gold), functionalized  with organic or biological
constituents can produce unique physical and chemical properties
that otherwise are not possible in single component
materials.\cite{hybrid} These materials have given rise to growing
interest in theoretical methods that can calculate the electronic
structure and transport properties of nanoscale
devices.\cite{refnano,refnano1,NR} A major problem with these
nanostructures is to optically control molecular self-organization
on metallic surface. J-aggregates of cyanine dyes make a fascinating
topic of research due to their outstanding optical
properties.\cite{kneipp} Spectroscopic peculiarities are the result
of exceptionally strong electronic interactions between the
transition dipole moments of the dyes that give rise to extended
exciton states after photo excitation.\cite{NR} The excitonic
optical spectrum depends on the details of the structural and
electronic arrangement of the dye molecule.\cite{OptSpect}

Adsorption of dyes to nanoparticles of noble metals presents a
special interest.\cite{PIC+Ag} Study of the interaction of an
adsorbed dye, in both its ground and excited states, with the energy
states of the conduction band of metal is also of interest and may
rely on theoretical calculations of the energy electronic structure
of the nanosystem made of {\em pseudoisocyanine (PIC) assembled on a
silver nanoparticle}. Moreover, the nature of coupling between
excitonic molecular J-aggregates and a metallic nanoparticle is
still not completely understood, since to the best of our knowledge
no attempts had been made to calculate the electronic structure of
Ag + PIC nanosystems from the "first principles" In addition, a
major problem of molecular engineering is to control molecular
self-organization.\cite{SelfOrg} This process is governed by the
formation of chemical bonds, e.g. covalent or hydrogen bonds. The
assembly process on nanoparticles is also affected by the molecule
--- substrate interactions. Therefore, understanding bonding between
the molecules and nanoparticle is crucial in order to be able to
choose appropriately the molecular and substrate materials for a
nanosystem design. On the theoretical side, the new computational
techniques allow one to predict the preferential geometry of the
nanostructure arrangements as well as the strength and chemical
bonds involved directly from the fundamental
quantum mechanical laws.

We demonstrate how {\em ab initio} calculations allow one to explain the
shape of the observed superstructures, to elucidate the role of
electronic structure and the molecule --- silver particle bonding
and to reveal details of the nanosystems not yet experimentally
accessible.

Quantum electronic-structure calculations allow us to understand the
macroscopic properties of complex polyatomic systems (specifically
organic PIC molecule assembled on a silver nanoparticle) in terms of
the microscopic states available to the electrons described by their
wave functions and the nanosystem electron density $\rho^{nano}(\bf
r)$. For these larger systems it is frequently of a greater interest
to know the change in the electronic structure associated with the
changes in the Kohn-Sham\cite{KS} effective potential $V_{eff}(\bf
r)$ and the electron density. Our work develops a theoretical
technique that makes the direct calculation of such changes
possible. This technique is based on the self-consistent
Green's function method\cite{FYKF} and standard density-functional
theory (DFT) in combination with the {\em local-density
approximation (LDA)},\cite{HK} which maps the many-electron
interaction problem onto a self-consistent treatment of
non-interacting quasi-particles moving in an effective potential
\begin{equation}\label{KS}
V_{eff}({\bf r})=V_{ext}({\bf r})+2\int\limits
\frac{\rho^{nano}({\bf r}')}{\mid {\bf r} - {\bf r}'\mid}d{\bf r}'
+ V_{xc}({\bf r}),
\end{equation}
where $V_{ext}({\bf r})$ is the external potential and $V_{xc}(\bf
r)$ is the local exchange-correlation potential.

\section{Formulation of the problem}

We present in this paper results of numerical calculations of the
electronic structure of nanosystem: PIC molecule + small Ag
particle, whose electronic structure is described by the Hamiltonian
\begin{equation}\label{Ham1}
H = H_{jellium} + H_{4d} + H_{PIC} + H_{PIC-Ag}^{p} + H_{PIC-Ag}^{r}
\end{equation}
The first term stands for the spherical jellium cluster Hamiltonian
(see below) made of $5s$-valence electrons of Ag-atom. The next term
in the Hamiltonian (\ref{Ham1}) describes the subsystem of Ag
$4d$-electrons embedded in the spherical jellium cluster. $H_{PIC}$
stands for the PIC molecule Hamiltonian. The last two terms of the
Hamiltonian (\ref{Ham1}) describe the $resonant (r)$ (due to the
presence of the 4$d$ state of Ag atoms) and $potential (p)$ (due to
the Ag 5$s$ states) scattering. This distinction between the two
types of scatterring by atomic potential is discussed in detail in
Ref. \onlinecite{KF} (see also \onlinecite{FYKF}).

The method is based on the separation of the Hamiltonian
(\ref{Ham1}) into the two parts
\begin{equation}\label{Ham2}
H = H_{LDA}^0 + \Delta H
\end{equation}
Here $H_{LDA}^0  = H_{jellium} + H_{4d}$ is the Hamiltonian of the {\em
reference} system, which includes an effective single-particle LDA
potential $V_{LDA}^0(\bf r)$. $\Delta H = H_{PIC} + H_{PIC-Ag}^{p} +
H_{PIC-Ag}^{r}$ is the Hamiltonian of the "difference" system
described by the potential $\Delta V(\bf r)$, which is the
difference between the self-consistent Kohn-Sham (\ref{KS})
effective potential $V_{eff}(\bf r)$ and $V_{LDA}^0(\bf r)$.

Four different computation schemes are used, each of which providing
mutually complementary information about the electronic spectra of
complex nanosystem. The first one is the spherical jellium cluster
model with the central $4d$ Ag atom embedded in the center of a
silver nanoparticle.\cite {KuF} The second method is based on the
Green's function approach. Then the physical effect shows up
exclusively in variations of the Kohn-Sham effective potential
$V_{eff}(\bf r)$ \cite{KS} and of the nanosystem electron density
$\rho^{nano}(\bf r)$. In this approach, we follow a two-step concept
and split the entire problem into two parts, where each of them is
far less complicated than the original one.\cite{Scheffler,Wachutka}
The whole nanosystem is decomposed into the reference  system
characterized by the ground-state electron density $\rho_{LDA}^0(\bf
r)$ and the difference system with the electron density
\begin{equation}\label{potential1}
\Delta \rho({\bf r})  =  \rho^{nano}({\bf r})  -  \rho_{LDA}^0({\bf
r}),
\end{equation}
which is the difference between the self-consistent electron density
$\rho^{nano}(\bf r)$ of the nanosystem and the ground-state electron
density $\rho_{LDA}^0(\bf r)$ corresponding to the Hamiltonian
\begin{equation}\label{Ham3}
H_{LDA}^0  =  -\nabla^2  +  V_{LDA}^0(\bf r)
\end{equation}
(atomic Rydberg units are used). Here $H_{LDA}^0  = H_{jellium} +
H_{4d}$. The complete set of electronic states of the reference
system is represented by the reference Green's operator
\begin{equation}\label{Greenf1}
G^0(z) = (z - H_{LDA}^0)^{-1}
\end{equation}
depending on the complex variable $z = \varepsilon + i\eta$
$(\eta\geq 0)$. The calculation of $G^0(z)$ along a properly chosen
contour in the complex energy plane defines step one of our
approach. As the second step, the Green's operator $G(z)$ of the
nanocluster-assembled system (and the charge density related to it)
is determined by solving the Dyson equation
\begin{equation}\label{Greenf2}
G(z) = G^0(z)  +   G^0(z)\Delta V G(z)
\end{equation}
in a self-consistent way. A detailed description of our approach to
solving Eq. (\ref{Greenf2}) is presented in Appendix.

The third semiempirical method ZINDO/S \cite {ZINDO} was used for the
calculation of electronic structure of a neutral PIC molecule. The
fourth DFT/B3LYP/6-31G \cite {GeomOpt} method was applied to the
equilibrium geometry calculation of PIC molecule by the standard
DFT-method.

We would like also to indicate here the limitations of our approach to
the treatment of Ag + PIC nanosystem based on the spherical jellium
model for the Ag nanoparticle. The previous structural research
\cite{BMF01} using method of the molecular dynamics shows that very
diverse nonspherical geometries of the nanoparticle are possible.
Relaxation of the PIC molecule and nanoparticle reconstruction in
response to the PIC molecule are also not taken into account.
Including all the processes into the calculation scheme can strongly
complicate the procedure and we leave it for our future research.

\section{Calculation of the electronic structure of Ag - nanoparticle}

First, we consider the electronic structure of silver nanoparticle.
Cluster calculations are traditionally employed in studies of
surface and bulk materials. They help us to understand how the
physical properties evolve from a free atom to a finite-size system.
In recent years, microclusters on the basis of Ag atoms attracted a
lot of interest, due to the growing technological significance of
nanosystems including Ag and Au nanoparticles.\cite{LCAO,Acherman} A
variety of theoretical models have been proposed for the
calculations of cluster electronic structure. However, the most
precise models allow one to obtain their electronic structure with a
small number of atoms only. For large atomic aggregations, these
methods cannot be applied successfully, and simpler models are
employed in this case. Thus, for the study of electronic properties
of sp-bonded metal clusters, a jellium model is used.\cite{Eckardt}
But the models based on the jellium approximation are not directly
suitable for the investigations of metal clusters containing atoms
with localized $d$ shells. As a consequence a model of an {\em Ag
atom embedded in the center of a spherical jellium cluster} is
applied for a description of a small Ag metallic nanoparticle
containing localized 4$d$ electrons. The DFT approach in the local
density approximation was used in computation.\cite{KuF,KuF1} Each
atom of the nanoparticle is mimicked by a single atom embedded in
the center of a jellium sphere with the $r_M$ radius, determined
according to the position of the real atom with respect to the
cluster surface. $r_M$ is then the shortcut between the atom and the
cluster surface.

The electronic structure of an Ag atom embedded in the jellium
sphere is obtained within the DFT framework from a self-consistent
solution of the Kohn-Sham equations (in atomic Rydberg
units)\cite{KS,KuF1}
\begin{equation}\label{Ham4}
[-\nabla^2 + V_{LDA}^0({\bf r})]\psi_{nl}({\bf r}) =
\varepsilon_{nl} \psi_{nl}({\bf r})
\end{equation}
where
\begin{equation}\label{potential2}
V_{LDA}^0(r) = -\frac{2Z^0}{r} + 2\int\limits \frac{\rho^{-}({\bf
r}') - \rho^{+}({\bf r}')}{\mid {\bf r} - {\bf r}'\mid}d{\bf r}'  +
V_{xc}(r)
\end{equation}
with the Vosko {\em {et al}} form\cite{Vosko} of the local
exchange-correlation potential $V_{xc}(r)$. Here $\varepsilon_{nl}$
and $\psi_{nl}$ are single electron energies and wavefunctions,
respectively; $Z^0$ is the nuclear charge of the Ag atom. The
electron density of the jellium cluster with an Ag atom is
\begin{equation}\label{density1}
\rho^{-}({\bf r}) = \sum_{nl} f_{nl}\mid\psi_{nl}({\bf r})\mid^2
\end{equation}
where the coefficients $f_{nl}$ are the occupation numbers of the
states with quantum numbers ${n,l}$, and the summation is over all
states of the {\em atom-in-jellium} nanoparticle. The radial
distribution of the positive jellium background is given by
\begin{equation}\label{density2}
\rho^{+}(r) = [3N_{val}(N_{at} - 1)/4\pi r_{M}^{3}]\Theta(r_{M} -
r),
\end{equation}
where $\Theta(x)$ is the unit step function, $N_{at}$ is the number
of atoms in the cluster (including the specific Ag atom), $N_{val}$
is the number of valence electrons in the Ag atom. The cluster
radius $r_{M}$ is found from the expression
\begin{equation}\label{radius1}
r_{M} = N_{at}^{1/3}r_{c},
\end{equation}
where $r_c$ is the Wigner-Seitz radius. We have used here $r_c =
3.02356$ a.u., $N_{at}=125$, and $N_{val}=1$. The numerical
integration of the Kohn-Sham equation for the Ag atom in a jellium
sphere is carried out by means of the Milne method.\cite{Miln} The
free Ag atom eigen-energies $\varepsilon_{nl}$ and wave-functions
$\psi_{nl}$ are calculated by means of the semirelativistic RATOM
program.\cite{RATOM} The self-consistency procedure for the
potential $V_{LDA}^0(r)$ is carried out in a mixed fashion. The
first two iterations use the arithmetic average scheme, which later
on is effectively substituted by the Aitken scheme.\cite{Aitken}

The electronic ground-state configurations of the jellium sphere
containing the central Ag atom were determined in the following way.
The rules for the energy level occupation separately in jellium and
in a free atom are well known. Clearly, insertion of an atom into
the jellium sphere center does not change the atom and jellium field
symmetry, and, consequently, the symmetry of their electronic states
is not changed either. One can suppose therefore that the number of
electron states with the same symmetry in the jellium sphere with
the central atom is equal to the sum of such symmetry states of the
jellium and the atom. The sequence of energy levels of {\em
atom-in-jellium} is obtained by solving the self-consistent
equations (\ref{Ham4}) - (\ref{density1}) for the different angular
$l$ and principal $n$ quantum numbers. The energy levels are
occupied in accordance with the Pauli principle. The highest levels
can be partially occupied.

We have computed the total energy of the jellium spheres with the Ag
atom in the center for various occupation numbers of the upper
levels. For the calculation of the total energy we employed the
equation
\begin{equation}\label{energy1}
E_{tot} = \sum_j^{occ} \varepsilon_j + \int \rho_{out}^{-}({\bf r})
\times
$$$$
\left[\varepsilon_{xc}^{out}({\bf r}) - V_{xc}^{in}({\bf r}) +
\int\limits\frac {\rho_{out}^{-}({\bf r'}) -
\mbox{2}\rho_{in}^{-}({\bf r'})} {\mid {\bf r} - {\bf r}'\mid} d{\bf
r}'\right] d{\bf r}.
\end{equation}
Here the indices {\em in} and {\em out} indicate the input and
output data of the latest self-consistency iteration, respectively;
$\varepsilon_{xc}^{out}({\bf r})$ is the exchange correlation energy
density of a homogeneous electron gas with the density
$\rho_{out}^{-}({\bf r})$ parameterized according to Vosko
{\em {et al}}.\cite{Vosko}

\begin{figure}[tbp!]
\begin{center}
\includegraphics[width=0.95\columnwidth]{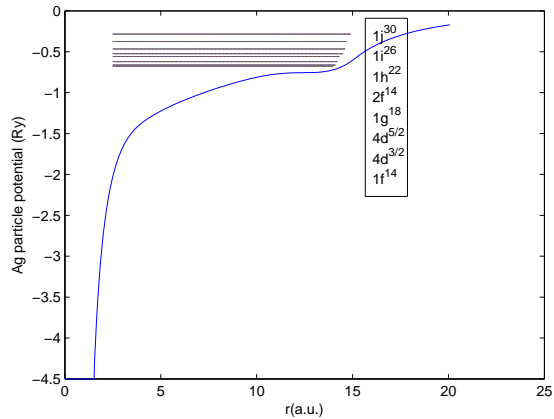}
\end{center}
\caption{(Color online) A self-consistent Ag-nanoparticle potential
of the jellium sphere containing the central Ag atom (the radial
distribution, in atomic units). The states of subshell $2d^{10}$ are
split into $4d^{5/2}$ and $4d^{3/2}$ levels due to the spin-orbit
interaction.} \label{Vparticle}
\end{figure}

The calculations have shown that the spherical jellium nanoparticle
with the central $4d$ Ag atom has the following energy spectrum. The
eight lowest energy levels are identical to those of the $4d$ atom
core; the rest of them are similar to the states in a spherical
potential well. In this paper the quantity $n = n_n + 1$ has been
considered as the principal quantum number; $n_n$ is the number of
nodes of the wave-function of the corresponding energy level.
Thus, the electronic configuration of the central Ag atom in the
jellium sphere is $1s^22 s^21 p^63s^22p^61d^{10}4 s^23p^6$. The
electronic states of the 125 atoms are arranged in the following
order: $2d^{10} 1f^{14} 1g^{18} 1h^{22} 2f^{14} 1i^{26} 1j^{30}$.
The self-consistent potential $V_{LDA}^0(r)$ (Eq.(\ref{potential2}))
with the electronic levels for an Ag atom embedded in a jellium
sphere is presented in Fig. \ref{Vparticle}.

An important characteristic of the Ag nanoparticles is their
ionization potential (IP). It has been shown in Ref.
\onlinecite{IP1} that a strong correlation exists between the
chemisorption reactivity of a small transition metal cluster and its
ionization threshold. Here we report the IP's for the Ag
nanoparticles containing up to 160 atoms. In these calculations we
simulate atoms of an Ag nanoparticle by Ag atoms embedded in the
center of jellium spheres of various sizes. The jellium sphere
radius $r_M$ is defined by the short cut between the atom and the
cluster surface.

The IP of an {\em atom-in-jellium} was obtained using the ground
state theory (LDA-method) by self-consistent calculation according
to Eq. (\ref{energy1}) and subtraction of the total energies of
neutral and ionic ground states. It was found that the ionization
thresholds of atoms of the Ag nanoparticles differ and depend on the
position of the atoms with respect to the cluster surface. The
lowest atomic ionization threshold has been chosen as the IP of the
Ag nanoparticle. For the neutral Ag atom we obtained
$IP_{atom}^{theor} = 7.96$eV and $IP_{atom}^{exp} = 7.576 $eV.
Instead of calculating the IP's from the data as the total energy
differences between the neutral cluster and ions, we use an
alternative method, based on the Slater transition state
approach\cite{TS} and Janak theorem \cite{DFTts} for the density
functional theory, which allow us to calculate the excitation energy
of adding (removing) an electron to (from) the system from (to) the
infinity. This scheme can be derived in the following way. The total
energy $E_{tot}$ difference between the final and initial states for
the process of electron addition to the one-electron state $j$ can
be calculated as an integral of total energy derivative with respect
to the occupancy $q_j$. This derivative, $\varepsilon_j = \partial
E/\partial q_j$, corresponds to the Kohn-Sham eigenvalue
\begin{equation}\label{energy2}
E(q_j=1) - E(q_j=0) = \int\limits_0^1 dq_j \varepsilon_j(q_j) \simeq
\varepsilon_j(0.5).
\end{equation}
Eq. (\ref{energy2}) becomes exact if the LDA eigenvalue
$\varepsilon_j(q_j)$ is a linear function of the occupancy, which is
usually true to within a good accuracy.

The vertical (or adiabatic) ionization potentials\cite{IPs} of the
clusters are evaluated from the {\em highest occupied molecular
orbital} (HOMO) energy of the neutral clusters. The calculated IP of
an Ag$_{125}$ nanoparticle is IP$^{theor} =4.73 $eV. When the number
of atoms in the nanoparticle increases the IP converges to the work
function $\Phi$ of the corresponding metallic half-space. The work
function of 4.46 eV was determined experimentally for
Ag(111).\cite{work} Our calculation yields the value of 4.73 eV for
IP$_{125}^{theor}$. The LDA HOMO-LUMO (lowest unoccupied molecular
orbital) gap between occupied ($\varepsilon_{i^{20}}$) and
unoccupied ($\varepsilon_{g^{18}}$) orbitals for the Ag nanoparticle
is $\Delta^{HOMO-LUMO} = \varepsilon_{1i^{20}} -
\varepsilon_{2g^{18}} = 1.58 $eV.

It is well-known that Ag aggregates are formed in a colloidal
aqueous solution of NaCl.\cite {PIC+Ag} Therefore studying
electronic properties of small metallic particles in different
dielectric matrices\cite{diel1,diel2,Lerme} presents a great
scientific and practical interest as well as those of free Ag
nanoparticles. When a metallic particle is placed in a dielectric
medium, polarization changes $\rho_{pol}(\bf r)$ are induced on the
particle surface, which produce the potential
\begin{equation}\label{potential3}
V_{pol}({\bf r}) = 2 \int \frac{\rho_{pol}({\bf r}')} {\mid {\bf r}
- {\bf r}'\mid} d{\bf r}'
\end{equation}
In Ref. \onlinecite{diel2}, we have shown that the ground state of a
metal nanoparticle embedded in a dielectric matrix can be described
by the self-consistent Kohn-Sham equations with the effective
potential
\begin{equation}\label{potential4}
V^{\varepsilon}({\bf r})= V({\bf r}) + V_{pol}({\bf r})
\end{equation}
where $V({\bf r})$ has the same form as the effective potential for
the nanoparticle in vacuum. Averaging $V_{pol}({\bf r})$ and
substituting the result into (\ref{potential4}) yields
\begin{displaymath}\label{potential5}
V^{\varepsilon}(r)=\left\{
\begin{array}{ll}
V(r) + \frac{1 - \epsilon}{\epsilon}V(r_M) & {r\leq r_M},\\
\\
V(r)/ \epsilon & {r \geq r_M}.
\end{array}\right.
\end{displaymath}
The calculations for the $Ag_{125}$ nanoparticle embedded in a
dielectric medium with the relative permeability $\epsilon = 61.1$
(50\% water solution of NaCl) have shown that with the increasing
dielectric permeability $\epsilon$ the potential profile near the
jellium edge becomes steeper. The bottom of the potential well
$V_{bottom}$ and the single electron energy levels $\varepsilon_i$
rise when the cluster is embedded in a dielectric matrix. It should
be emphasized that the oscillations\cite{diel2} in the electronic
density of jellium cluster in vacuum are suppressed in the
dielectric media. The amount of the electronic charge beyond the
jellium edge (electronic "spill out") increases with the increasing
$\epsilon$. This is caused by a positive shift of the electronic
eigenenergies of Ag nanoparticles in a dielectric medium and an
extension of the corresponding wave functions.

The static dipole polarizability\cite{ZS} is also well-known to be
intimately related to the electronic structure of a
nanocluster.\cite{diel2} We discuss here the photo-response of an
isolated Ag nanoparticle to an external electromagnetic field in
terms of the frequency-dependent polarizability\cite {Puska}
\begin{equation}\label{polarizability}
\alpha(\omega) = - \frac{8\pi}{3} \int \limits_{0}^{\infty} dr' r'^3
\delta\rho(r',\omega)
\end{equation}
where $\delta\rho(r',\omega)$ is the change in the charge density.
The calculation of the static dipole polarizability
$\alpha(\omega=0)$ for an Ag nanoparticle is carried out within the
time-dependent local-density approximation (TDLDA)\cite{ZS} using
the self-consistent solution of the set of equations
\begin{equation}\label{density3}
\delta\rho({\bf r},\omega)=\int \chi_0({\bf r},{\bf r}';\omega)
\delta V({\bf r}') d{\bf r}',
\end{equation}

\begin{equation}\label{density4}
\delta V({\bf r},\omega)= V_{ext}({\bf r},\omega)+ V_{ind}({\bf
r},\omega),
\end{equation}
and
\begin{equation}\label{potential6}
V_{ind}({\bf r},\omega) = \rm 2 \int \frac{\delta\rho({\bf
r}',\omega)} {\mid {\bf r} - {\bf r}'\mid} d{\bf r}' +
\frac{\partial V_{xc}({\bf r})} {\partial\rho({\bf r})}
\delta\rho({\bf r},\omega).
\end{equation}
Here $V_{ext}({\bf r},\omega)$ and $V_{ind}({\bf r},\omega)$ are the
external field with the frequency $\omega$ and the induced field,
respectively; $\chi_{0}({\bf r},{\bf r}';\omega)$ is the
susceptibility function in the independent-particle approximation.
For $r_M =15.1178$ a.u. (Ag$_{125}$ cluster) $\alpha(\omega=0) =
4.82 \AA^3/$ atom. The bulk atomic polarizability is 4.33
$\AA^3$/atom.\cite{polar}

\section{Results and discussion}

We present here a theoretical approach to the organic molecules
interacting with silver in terms of numerically solvable DFT-models.
In practice, these models apply to nanosystems (PIC-molecule on
Ag-particle) and provide an understanding of doping silver
nanoparticles by PIC adsorbates.
\begin{figure}[tbp!]
\begin{center}
\includegraphics[width=0.45\columnwidth]{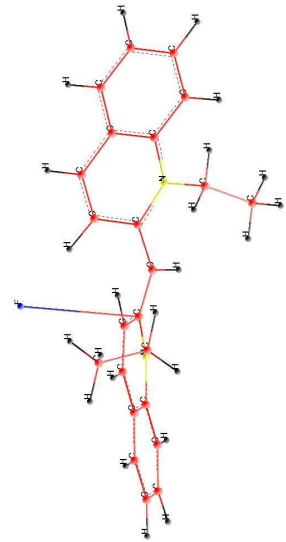}
\end{center}
\caption{(Color online) Configuration of the PIC molecule calculated
by DFT/B3LYP/6-31G-method: H(black) stand for the hydrogen atoms;
C(red) stands for the carbon atoms; N(yellow) stands for the
nitrogen atoms; F(blue) stands for the fluorine atom.}
\label{fig.11}
\end{figure}
The chemical structure of PIC molecule is exhibited in Fig.
\ref{fig.11}. The density-functional calculations are performed in
order to determine the equilibrium structure of these PIC molecules
by the DFT/B3LYP/6-31G-method.\cite{GeomOpt} The molecule geometries
are optimized using the B3LYP-exchange-correlation energy functional
and potential. The minimum basis set is capable of producing reasonable
results is 6-31G. Since the geometry of PIC within the
Ag+PIC-nanosystem is undoubtedly disturbed as compared to the PIC
monomer and possibly resembles a PIC aggregate, these calculations
were performed as follows. First, the equilibrium geometry of the
tetramer (PIC:Cl)$_4$ was found. After that, the averaged geometry
of the monomer (PIC:Cl) was obtained by averaging over four
molecules of the tetramer. The search of the equilibrium geometry of
tetramer was performed by standard density functional method
DFT/B3LYP/6-31G using GAMESS program package.\cite {GeomOpt} The
electronic structure of a neutral PIC molecule was calculated by
ZINDO/S-method.\cite {ZINDO}

It is assumed that the PIC molecule is adsorbed on the spherical
surface of an Ag nanoparticle, setting down over the sphere so that
its central carbon atom (the molecule center of gravity) is
positioned in the point (2.223,0.680,0.705)\AA with respect to the
center of Ag nanoparticle. In the beginning we present the results
of calculation of the electronic structure of an Ag$_{125}$
nanoparticle obtained by the model of an {\em atom in the center of
a spherical jellium sphere} and DFT in the local density
approximation. Fig. \ref{Vparticle} presents the energy level
structure of the Ag jellium nanoparticle with 134 valence electrons
(the Ag atom core electrons orbitals lie below
$\varepsilon_{3p^{3/2}}$ = - 4.27 Ry ).
\begin{figure}[tbp!]
\begin{center}
\includegraphics[width=1.0\columnwidth]{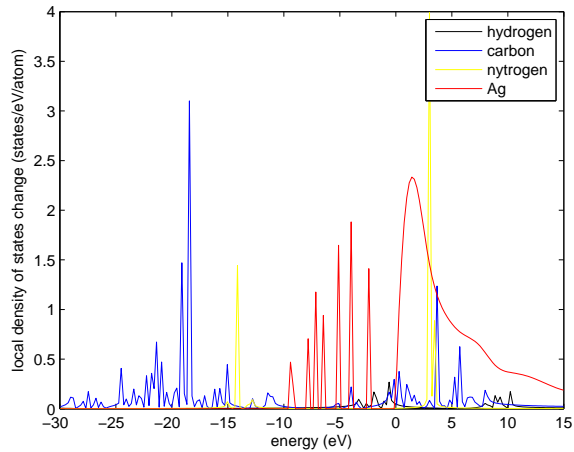}
\end{center}
\caption{(Color online) Density of states and Kohn-Sham level
structure for nanosystem $Ag+PIC$. } \label{fig.5}
\end{figure}

Fig.\ref{fig.5} shows the Ag bound states by the red line. The peak
heights are proportional to the degeneracy. The continuum density of
states $\Delta n(\varepsilon)$ is given for the positive energies
corresponding to the delocalized (unoccupied) states by Eq.
(\ref{density5}). Fig. \ref{fig.5} exhibits the contribution of
potential scattering [Eq.(\ref{pole})] to the formation of Ag
related bound states ($\varepsilon_{2g}$ =-1.34 $eV$) and fall in
the broad range $\Delta^{HOMO-LUMO}$. This level is obtained from
Eq.(\ref{pole}) with $\widetilde{\cal M}$ substituted for $Q^{ -1}$
[see Eq.(\ref{Dyson})]. The states $2d^{10}(4d^{3/2}+4d^{5/2})
1f^{14}1g^{18}1h^{22}2f^{14}1i^{26}1j^{30}$ in the occupied part of
the spectrum are the Ag resonances in the band (red line in
Fig.(\ref{fig.5}) and lie in the interval (-10,-5) eV. The resonant
$\varepsilon_{2g}$ level [see Fig.\ref{fig.5}] arises at the energy
0.24 eV below the LUMO-level. Both the Ag resonances and
$\varepsilon_{2g}$ states are found in the solution of
Eq.(\ref{pole}) with the full self-energy $\widetilde{\cal M}(z)$.

The calculation of Ag$_{125}$ particle embedded in a dielectric
medium with relative dielectric permeability $\varepsilon = 61.1$
(50\% water solution of NaCl) have shown that the electronic "spill
out" increases in the dielectric medium as compared to vacuum and is
equal for this case to 21.6 electrons.

The 26.14 eV width band of Ag+PIC nanostructure is formed by
strongly hybridized C(s,p) (blue line), N(s,p) (yellow line), H(s)
(black line) and Ag (red line) states. The unoccupied states lie
above $\varepsilon^{LUMO} = -1.58$ eV. Furthermore, we find the
acceptor-like $\varepsilon_{2g}$ states almost entirely localized in
the adsorbate PIC molecule and hybridized with the hydrogen 1$s$
levels. As discussed in Ref. \onlinecite{hyd}, the interaction
between the adsorbate and the transition metal surface can be
described as a two-state problem (adsorbate state and the d-band)
leading to formation of bonding and anti-bonding states. Thus, an
upshift of the d-states should increase the adsorbate-metal
interaction, since it would lead to the formation of an anti-bonding
orbital closer to the Fermi level. Strong features appearing between
-5 and -10 eV represent formation of a bonding orbital through the
interaction of hydrogen 1s state with the metal d-band, and this
formation is typical for all transition metals. The bonding Ag 2d-1s
states lie around -9 eV. The anti-bonding orbital lies around
$\varepsilon^{LUMO}$ level and is hybridized with the resonance
$\varepsilon_{2g}$ states.

\begin{figure}[tbp!]
\begin{center}
\includegraphics[width=0.95\columnwidth]{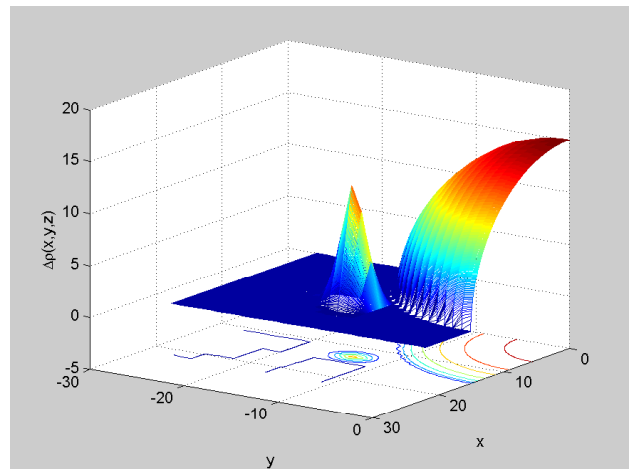}
\end{center}
\caption{(Color online) Difference electron-density charge for
$PIC$- molecule adsorbed on an Ag-nanoparticle. The contours are in
units of $10^{-5}e/\Omega_A$.} \label{rhoSphere}
\end{figure}

We will interpret the density of states (see figure \ref{fig.5}) of
nanosystem Ag+PIC species basing on their bonding properties
obtained from the Green's function calculations and simple molecular
orbital considerations. For a more detailed study of the chemical
bonding in these nanosystem we have calculated the electron-density
change $\Delta\rho(\bf r)$ (see figure \ref{rhoSphere}).

The electronic structure obtained for the Ag+PIC nanosystem has a
terminal Ag---C bond. Similarly to all transition
metals\cite{CCbound} the C---C bond is known to be much stronger
than the Ag---C bond. Thus, it is not energetically favorable for Ag
to enter the C chain, since this will break the stronger C---C bonds
and form weaker Ag---C bonds. The Green's function calculations
indicate that the excited Ag atoms in the nanosystem are involved in
the chemical bonding. Simple molecular orbital considerations
suggest that the C atom directly bonded to an Ag atom must be in an
$sp$-hybridized state in order to more effectively interact with the
Ag particle orbitals. And the Ag $2d$ orbitals can interact with the
other $p$ orbitals of the C atoms to form additional bonds. Thus,
our calculations show that the Ag(d,s) --- C(s,p) bond is of a
considerable importance for understanding the nature of chemical
bonding of Ag+PIC nanosystem.

As shown in Fig.\ref{fig.5}, all Ag states are situated between
the C-states. From $\Delta n(\varepsilon)$ distribution as well as
from its partial components $\Delta n_s(\varepsilon)$, $\Delta
n_p(\varepsilon)$ and $\Delta n_d(\varepsilon)$, one can conclude,
that $d-s,p$ resonance is observed in (Ag+PIC)-nanosystem, was
revealed in the compounds of metals, containing the filled d-shells,
with non-metals.\cite{AgCl} In this case,the Ag 5s-type states
($1f^{14} 1g^{18} 1h^{22} 2f^{14} 1i^{26} 1j^{30}$) form $\sigma$
bonding and $\sigma^*$ antibonding orbitals with the C $sp$-
hybridized orbitals. At the same time, the 2d$\pi$ orbitals in Ag
can interact with the C---C $\pi$ orbitals. The resonance of the
HOMO $\varepsilon_{2g}$ with the hydrogen $1s$-level
($\sigma^*$-symmetry) is the principle interaction in the
Ag$5s$---H$1s$ bond. This bond plays the crucial part in determining
the chemisorption reactivity.\cite{reactiv}

There are two factors that affect this orbital interaction, first
the HOMO energy, and second and more importantly, the symmetry of
the HOMO. It follows from Ref. \onlinecite{reactiv} that the
d-symmetry HOMO would have lead to a high reactivity. However, the
$\varepsilon_{2g}$ HOMO is mainly of 5$s$ character and is not
symmetry matched with the H $\sigma^*$ orbital, hence it results in
an extremely low reactivity. Thus, the $s$-type HOMO of the Ag+PIC
nanosystem effectively provides a {\em shielding effect} to protect
the nanosystem from being attacked by the hydrogen atoms.

Fig. \ref{rhoSphere} shows the charge density difference of an Ag +
PIC nanosystem
$$
\Delta\rho({\bf r} ) = \rho^{Ag+PIC}({\bf r}) - \rho^{Ag}({\bf r}) -
\sum_a \rho_a^{atom}.
$$
Here, $\rho^{Ag+PIC}({\bf r})$ is the self-consistent electron
density of the Ag + PIC nanosystem; $\rho_a^{atom}$ is the electron
density of the free atoms in the PIC-molecule; $\rho^{Ag}({\bf r})$
is the density in reference system. The Ag sphere with radius 15.12
a.u. is also schematically shown. The electron-density charge is
close to $z=1.33 $ a.u. Fig. \ref{rhoSphere} reveals that the
electron density change $\Delta\rho(\bf r )$ is strongly localized.
We see also that the adsorbed PIC-molecule causes an accumulation of
the reference charge on Ag-metal side due to the three carbon atoms
with the coordinates (14.21, -13.71,1.33) a.u., (12.13,-12.03,0.99)
a.u., (13.93,-16.37,1.31) a.u. and one hydrogen atom with the
co-ordinates (15.25,-17.37,2.26) a.u. Based on the chemical bonding
model of the carbon and hydrogen valences, the chemical structure of
interaction between the Ag and (C,H) atoms of PIC-molecule has the
form
$$ H - Ag - C = C - C - H$$
The major part of electron density in $C=C$ and $C=N$ bonds is
localized at the bottom of the band and have the $sp$ and $sp^2$
type covalence of the chemical bonding. At the same time, the upper
part of the occupied states contain mainly the bonding electrons
which are concentrated in Ag---H and Ag---C bonds and have
($d,s$)--($s,p$) resonance type of bonding with the maximum
displacement towards the Ag nanoparticle.

\section{Concluding remarks}

Carrying out a numerical solution of the Dyson equation
(\ref{Dyson}) in the Kohn-Sham density-functional methodology we
determined the electronic and chemical structure of Ag+PIC
nanosystems. The calculation of Ag$_{125}$-particle embedded in a
dielectric medium with the relative permeability $\epsilon=61.1$
(50\% water solution of NaCl) shows that the amount of the
electronic charge beyond the jellium edge (electronic "spill out")
increases in the dielectric medium as compared to the vacuum and
equals in this case 21.6 electrons. The resonant
$\varepsilon_{2g}$-level [see Fig.\ref{fig.5}] arises at the energy
0.24eV under the LUMO-level. Both the Ag - resonances and
$\varepsilon_{2g}$ states are found as solutions of Eq.(\ref{pole})
with the full self-energy $\widetilde{\cal M}(z)$. The electronic
structure obtained for the Ag+PIC nanosystem has a terminal Ag---C
bond.

Three C atoms and one H atom take part in the adsorption of the PIC
molecule on the Ag nanoparticle and accumulation $\Delta \rho$ of
charge takes place. The hybridized Ag --- H and Ag --- C bonds are
of the ($d,s$)-($s,p$) resonance type  with the maximal displacement
of charge towards the Ag nanoparticle.

{\bf Acknowledgment} Authors are indebted to K.Kikoin for detailed
discussions. This work was supported by the Russia-Israel Scientific
Research Cooperation, grant N 3-5802 (B.F. and V.M.) and the
Israel-US Binational Science Foundation (B.F.), grant N 2008282.

\appendix

\section{Green's-function technique to calculate the electronic
properties of organic admolecule on metal cluster}

Our approach is based on the general concept formulated in Ref.
\onlinecite{Green-func}. It makes it possible to calculate the
electronic structure of metallic substrates and to study the
behavior of a compact cluster adsorbed on a nanoparticle. The
principal theoretical tool is the scattering theory formalism, which
considers perturbation of a metal substrate single electron
potential by a spatially compact cluster, which is itself compact by
virtue of screening. It can therefore be treated as a localized
scattering potential for Ag electrons of the substrate. In order to
obtain the electronic structure of admolecule on an Ag nanoparticle
a "matrix" scattering approach is adopted. Our objective is to
construct the Green's function matrix $G_{\mu \nu}(z)$ for the
perturbed system, e.g., silver nanoparticle with organic
PIC-molecule.

The theoretical analysis providing us with a link between the
nanosystem and the related simple systems (Ag nanoparticle and
PIC-molecule) is based on the Dyson equation (\ref{Greenf2}). As is
well known, the original problem of solving a linear differential
equation can be mapped onto the solution of a matrix equation of
infinite dimension by expanding the wave functions in terms of a
linear combination of properly chosen orthonormal functions, e.g.
atomic orbitals (LCAO's) that are used here. In most cases of the
LCAO cluster calculations Slater-type orbitals strongly facilitate
the numerical calculation of overlap integrals.

In practical calculations we construct the Green's function matrix
$G_{\mu \nu}(z)$ (\ref{Greenf1}) using a finite set of $N_b$ basis
functions $\chi_{\mu}(\bf r)$. The basis set only needs to cover the
real space, within which $\Delta \rho({\bf r})$ is localized. We
will denote this region as box $A$ of volume $\Omega_A$. In the
present implementation of our method the $\chi_{nlm}(\bf r)$ are the
Kohn-Sham orbitals
\begin{equation}\label{orbital1}
\chi_{nlm}({\bf r}) = \psi_{nl}(\mid{\bf r - R}_{\alpha}\mid)
Y_{lm}(\theta_{{\bf r - \bf R}_{\alpha}},\phi_{{\bf r-\bf
R}_{\alpha}})
\end{equation}
placed at appropriately chosen positions $\bf R_\alpha$ in the PIC
assembled molecule. Here we choose the basis of atomic functions so
that $\mu \equiv nml$. $\psi_{nlm}({\bf r})$ are the Kohn-Sham
radial wave functions obtained with the help of the potential
$\Delta V({\bf r})$, which in turn is calculated by means of the
functions $\psi_{nlm}({\bf r})$. The selfconsistency iterations are
repeated until a desired convergence is achieved.
$Y_{lm}(\theta,\phi)$ are the spherical harmonics centered at ${\bf
R}_\alpha$.

Writing the Dyson equation
\begin{equation}\label{Greenf3}
\textsf G(z) = {\textsf G}^0(z)  +  \Delta\textsf G(z)
\end{equation}
we see that only the difference operator $\Delta \textsf G(z)$ (and
the difference electron charge density $\Delta \rho (\bf r)$ related
to it) need to be be actually calculated. This difference operator
has the form
$$
\Delta\textsf{G}(z) = \left[\left({\textsf I} - \widetilde {
{\textsf G}}^0(z)\cdot \Delta{\cal V} \cdot ({\textsf
L}^{-1})^\dagger \right)^{-1} - {\textsf I} \right] \widetilde{
\textsf G}^0(z)
$$
where
$$
\Delta{\cal V}_{\mu\nu} = \int_{\Omega_{A}} \chi^*_\mu({\bf r})
\Delta V[\rho({\bf r})] \chi_\nu({\bf r}) d{\bf r}
$$
and ${\textsf I}$ is a unit matrix. Here the factor $\textsf L^{-1}$
ensues from the assumption that the basis set $\chi_{\mu}(\bf r)$
was used in the Cholesky decomposition ${\textsf S}= {\textsf
L}\cdot {\textsf L}^\dagger$ for the overlap matrix
$$
S_{\mu\nu} = \int_{\Omega_{A}} \chi_\mu^*({\bf r}) \chi_\nu({\bf
r}) d{\bf r}
$$
in order to obtain the orthonormal basis.

The density variation is calculated using the equation
\begin{equation}\label{deltarho}
\Delta \rho({\bf r}) = \mbox{Im} \sum_{\mu=1}^{N_b}
\sum_{\nu=1}^{N_b} \widetilde{\Delta\rho}_{\mu\nu} \chi_\mu({\bf r})
\chi^*_\nu({\bf r})
\end{equation}
where
$$
\widetilde{\textsf G}^0_{\mu\nu}(z) = \left(({\textsf
L}^{-1})^\dagger {\textsf G^0(z)} {\textsf L}^{-1}\right)_{\mu\nu},
$$
$$
\widetilde{\Delta\rho}_{\mu\nu} = \left(({\textsf L}^{-1})^\dagger
{\Delta\rho} {\textsf L}^{-1}\right)_{\mu\nu}
$$
and
\begin{equation}\label{deltarho_1}
\Delta\rho = - \frac{1}{\pi}
\int_{\varepsilon_b}^{\varepsilon_{HOMO}} \Delta\textsf{G}(z) dz.
\end{equation}
The lower integration limit $\varepsilon_b$ is chosen in such a way
as to include all the relevant Ag and molecule states,
$\varepsilon_{HOMO}$ is the HOMO energy. To compute the integral
(\ref{deltarho_1}), we introduce a contour $C$ in the complex plane
$z$ enclosing all the poles of the Green's function up to the
highest occupied molecular orbitals energy in the charge density
integration.

Our computational scheme is based on the spherically symmetric Ag
metallic nanoparticle Green's function
\begin{equation}\label{Greenf4}
G^0({\bf r,\bf r}';\varepsilon) = \sum_{lm} Y_{lm}(\theta,\phi)
G_l^{0}(r,r';\varepsilon) Y_{lm}^{*}(\theta',\phi'),
\end{equation}
decomposed in terms of the spherical harmonics.\cite{ZS} Here
\begin{equation}\label{Greenf5}
G_{l}^{0}(r,r';\varepsilon) = \frac{R_l(r_{<},\varepsilon)
N_l(r_{>},\varepsilon)}{r^2 W_l(\varepsilon)},
\end{equation}
with $r_{<}=min(r,r^{'}), r_{>}=max(r,r^{'})$.
$R_l(r_{<},\varepsilon)$ and $N_l(r_{>},\varepsilon)$ are the
regular and nonregular solutions of the radial Kohn-Sham equations
with the potential $V^0(r)$ and energy $\varepsilon$.
$W_l(\varepsilon)$ is the Wronskian of the functions $R_l$ and
$N_l$. We obtain $R_l$ and $N_l$ by integrating the radial Kohn-Sham
equation, using the asymptotic behavior of these functions at the
origin and infinity. As is generally known, the regular solution
$R_l$ behaves asymptotically as $r^{l+1}$ at $r \rightarrow 0$. When
$r \rightarrow \infty$, the nonregular solution $N_l$ must be an
outgoing wave for the continuum-energy region($\varepsilon > 0$),
and it exponentially decreases for the bound-state region
($\varepsilon < 0$).

The Green's function of the nanoparticle is projected onto the
localized basis
\begin{equation}\label{Greenf6}
G_{\mu\nu}^0(\varepsilon) = \sum_{lm} \langle\mu\vert \Theta_A\vert
R_l (r_{<},\varepsilon)\rangle\langle N_l( r_{>},\varepsilon)
\vert\Theta_A\vert\nu\rangle{/}W_l(\varepsilon)
\end{equation}
The following notations has been used above:
$$
\langle\mu\vert\Theta_A\vert R_l(r_{<},\varepsilon)\rangle =
\int_{\Omega_{A}} \chi_\mu^*({\bf r-R}_\alpha)R_l(r_{<},\varepsilon)
d{\bf r},
$$
$$
\langle N_l(r_{>},\varepsilon)\vert\Theta_A\vert\nu\rangle=
\langle\nu\vert\Theta_A\vert N_l(r_{>},\varepsilon)\rangle^*,
$$
$$
\langle\nu\vert\Theta_A\vert N_l( r_{>},\varepsilon)\rangle =
\int_{\Omega_A} \chi_\nu^*({\bf r-R}_\beta)N_l (r_>,\varepsilon)
d{\bf r},
$$
$$
\Theta_{A}({\bf r}) = \left\{
\begin{array}{ll}
1, & {\bf r} \in \Omega_{A}\ -\ \mbox{volume of the}\\
&\ \mbox{box
region A}\\
&\\
0, &\ \mbox{otherwise},
\end{array} \right.
$$

Here we use the single-center method to compute all integrals in the
Slater-type orbital basis.\cite{SlaterOrbital} The Slater orbital
centered in a point, defined by its location vector $\bf
R_{\alpha}$, is usually
\begin{equation}\label{orbital2}
\chi_{nlm}({\bf r}) = N_{nl}\mid{\bf r} - {\bf
R}_{\alpha}\mid^{n-l-1} \exp(-\zeta\mid{\bf r} - {\bf
R}_{\alpha}\mid)
$$
$$
\times Y_{lm}(\theta_{{\bf r} - {\bf R}_{\alpha}},\phi_{{\bf r} -
{\bf R}_{\alpha}})
\end{equation}
The radial part of the Slater orbital is expanded over the Barnett-
Coulson/L\"owdin function (BCLF),\cite {BCLF}
$$
\mid{\bf r} - {\bf R}_{\alpha}\mid^{ n-l-1} \exp(-\zeta\mid{\bf r} -
{\bf R}_{\alpha}\mid)=
$$
\begin{equation}
\frac{1}{\sqrt{R_{\alpha}r}} \sum_{\lambda}(2 \lambda + 1) {\cal
A}_{\lambda + 1/2}^{ n-l} (\zeta,R_{\alpha},r)P_\lambda
\left(\frac{{\bf R}_{\alpha}\cdot{\bf r}}{ R_{\alpha}r}\right)
\end{equation}
where $P_n(z)$ are the Legendre polynomials of degree $n$ and ${\cal
A}_{\lambda+1/2}^{n-l} (\zeta,R_{\alpha},r)$ are the BCLF's defined
by the recursion
$$
{\cal A}_{\lambda+1/2}^{n} (\zeta,R_{\alpha},r)= - \frac{\partial}{
\partial\zeta} {\cal A}_{\lambda+1/2}^{n-l} (\zeta,R_{\alpha},r)
$$
with
$$
{\cal A}_{\lambda+1/2}^{0} (\zeta,R_{\alpha},r)= {\bf
I}_{\lambda+1/2}(\zeta\rho_<) {\bf K}_{\lambda+1/2}(\zeta\rho_>)
$$
where ${\bf I}_{\lambda+1/2}(z)$ and ${\bf K}_{\lambda+1/2}(z)$ are
the modified Bessel functions of the first and second kind; the
variables $\rho_<$ and $\rho_>$ stand for the $min(R_{\alpha},r)$
and $max(R_{\alpha},r)$ respectively. In the present implementation
of our method the Kohn-Sham orbitals are placed at appropriately
chosen positions $\bf R_{\alpha}$.

Information on the Ag+PIC adsorption can be produced from the Dyson
equation (\ref{Greenf2}) that describes the interaction between the
free molecule and silver particle. In this Dyson equation the part
of the unperturbed Green's function is played by the {\em Augmented
Green Function}
\begin{equation}\label{augmgreen}
{\sf{G}}^{NS}_{Ag+PIC}(z) = \left(
\begin{array}{cc}
G^0_{PIC}(z) & 0 \\
&\\
0 & G^0_{Ag}(z)
\end{array}
\right)
\end{equation}

In order to describe the molecule --- silver nanoparticle
interaction we use the approach known in the theory of transition
metal impurities in semiconductors
\cite{KF,Haldane,FK86,And61,Fleurov1}. The terms in the model
Hamiltonian (\ref{Ham1}) for Ag+PIC nanosystem are written as
$$
H_{PIC-Ag}^{r} = \sum_{ki_a}M_{ki_a}c_{k}^\dagger d_{i_a}  + h.c.
$$
and
$$
H_{PIC-Ag}^{p} = \sum_{kk'}W_{kk'}c_{k}^\dagger c_{k'}
$$
Here, $c_{k}$ ($d_{i_a}$) and $c_{k}^\dagger$ ($d_{i_a}^\dagger$)
denote the usual fermionic annihilation and creation operators,
respectively, which are labeled by the indexes, $k$ and ${i_a}$,
containing site, orbital and spin degrees of freedom; $M_{ki_a}$ is
the $s,d-{i_a}$-hybridization matrix element, and $W_{kk'}$ is the
matrix element for the Ag particle short-range potential scattering
(the subscript "a" refers to the a-th adatom in the PIC-admolecule
and ${i_a}$ stands for the corresponding electronic state and atom
site). The Dyson equation reads
\begin{equation}\label{Dyson}
G_{i_ai_a}(z)=G_{i_ai_a}^{NS}(z)[1+\widetilde{\cal M}_{i_a}(z)
G_{i_ai_a}(z)]
\end{equation}
where
$$
G_{i_ai_a}^{NS}(z) = \frac{1}{z - \varepsilon_{i_a}-\Delta V_{i_ai_a}}
$$
$$
\widetilde{\cal M}_{i_a}(z)= {\cal M}_{i_a}( z)Q^{ -1}( z)
$$
Off-diagonal elements even for the nearest neighbors are up to two
orders of magnitude smaller than the diagonal ones, and are
neglected.

The PIC electron levels $\varepsilon_{i_a}$ are found
self-consistently as solutions of the Kohn-Sham equations for the
PIC related orbitals in the nanosystem environment. The self-energy
${\cal M}_{i_a}(z)$ contains two contributions. The term
\begin{equation}\label{massoper}
{\cal M}_{i_a}(z) = \sum_k\frac{|M_{k{i_a}}|^2}{z - \varepsilon_k}
\end{equation}
describes the hybridization between the $s,d$-PIC-orbitals
($\psi_{i_a}$) and the Ag electrons ($\psi_k$) with the matrix
element
\begin{equation}\label{hyb}
M_{ki_a}= \int_{\Omega_{A}} \psi^{*}_{k}({\bf r}) \Delta V({\bf r})
\psi_{i_a}({\bf r-R}_{\alpha}) d{{\bf r}}.
\end{equation}
The factor
\begin{equation}\label{Kosla}
Q(z) = 1 - \Delta V_{pot} G_{Ag}^0(z).
\end{equation}
in Eq. (\ref{Dyson}) describes the short-range potential scattering,
where
\begin{equation}\label{sub}
\Delta V_{pot} = \sum_{kk'}\int\psi^{*}_k({\bf r}) \Delta V({\bf r}
) \psi_{ k'}({\bf r}) d{\bf r}.
\end{equation}
and
\begin{equation}\label{greenol}
G_{Ag}^0(z) = \sum_k\langle k|(z - H_{LDA}^0)^{-1}|k\rangle =
\sum_k\frac{1}{z - \varepsilon_k}
\end{equation}
is the single-site lattice Green's function for the electrons in the
Ag host cluster described by the Hamiltonian $H_{LDA}^0$.

The Ag electron levels $\varepsilon_{k}$ are found self-consistently
as solutions of the Kohn-Sham equations for Ag-related orbitals in
the nanosystem environment. The Green's function (\ref{Dyson})
describes the hybridization between the PIC-electron orbitals and
the electrons in the silver host cluster, where the Ag electrons are
influenced by the potential scattering due to $\Delta V$. If this
scattering is strong enough, it results in splitting off of
localized levels from the top of the band. This effect is also taken
into account in Eq. (\ref{Dyson}): the positions of the
corresponding levels before the hybridization are determined by the
zeros of the function $Q(z)$ in the energy gap $\Delta^{HOMO-LUMO}$
of the Ag+PIC nanosystem, which arises in the electronic structure
due to the potential scattering only. As a result the equation for
the deep level energy determined as a pole of the Green's function
[Eq. (\ref{Dyson})] within the framework of the LDA technique reads
\begin{equation}\label{pole}
z - \varepsilon_{i_a} - \Delta V_{i_ai_a}^{LDA} = \widetilde{{\cal
M}}_{ i_a}( z)
\end{equation}
It takes into account the resonance part of the scattering amplitude
in the $i_a$ (PIC) channel and its mixing with the potential
scattering states arising in the $k$ (Ag) channel.

The imaginary part of the Green's function yields the spatial and
energy electron distribution
$$
\rho({\bf r},\varepsilon) = - \frac{2}{\pi}\mbox{Im} G(\bf
r,r;\varepsilon).
$$
Integrating over the energy we get the charge density distribution,
whereas the local density of states, e.g., in the cell $\Omega_{A}$,
reads
$$
n_{loc}(\varepsilon) = \int_{\Omega_A}\rho({\bf r},\varepsilon)
d{\bf r}
$$
The change of the density of states is
$$
\Delta n(\varepsilon) = \mbox{Tr}( G_{Ag+PIC}) - \mbox{Tr}(
G_{Ag}^0)
$$
or after straightforward calculations
\begin{equation}\label{density5}
\Delta n(\varepsilon)=
$$$$
\frac{2}{\pi} \mbox{Im} \sum_{i_a} \frac{d}{d\varepsilon} \mbox{ln}
[(z - \varepsilon_{i_a} - \Delta V_{i_ai_a}) Q(\varepsilon) - {\cal
M}_{ i_a}( z)]
\end{equation}

The problem is treated self-consistently, starting with the trial
set of LCAO Slater-type functions. The difference potential in the
zero approximation is a sum of the atomic potentials of the
nanosystem. The selfconsistency procedure for $\Delta V({\bf r})$ is
carried out in a mixed fashion. The first two iterations use the
arithmetic average scheme, which later on is effectively substituted
by the Aitken scheme.\cite{Aitken} Just seven iterations produce a
$10^{-4}$ Ry selfconsistency.

\end{document}